\documentclass[11pt]{article}

\usepackage{amsfonts,latexsym,eucal,color,graphicx,epsfig,amssymb,amsmath}

\newcommand{\be}{\begin{eqnarray}}
\newcommand{\ee}{\end{eqnarray}}
\newcommand{\beq}{\begin{eqnarray}}
\newcommand{\eeq}{\end{eqnarray}}

\newcommand{\dalm}{\kern1pt\vbox{\hrule height 0.9pt\hbox{\vrule width 0.9pt\hskip 2.5pt\vbox{\vskip 5.5pt}\hskip 3pt\vrule width 0.3pt}\hrule height 0.3pt}\kern1pt}

\newcommand{\tr}{\textrm{tr}}

\begin{document}


\begin{titlepage}

\vskip 2.5cm
\begin{center}
{\bf\LARGE{Generalized entropy and higher derivative Gravity
}}
\vskip 1.5cm
{\bf 
Joan Camps
}
\vskip 0.5cm
\medskip
\textit{DAMTP, Cambridge University}\\
\textit{Wilberforce Road, Cambridge CB3 0WA, United Kingdom}\\

\vskip .2 in
\texttt{J.Camps@damtp.cam.ac.uk}

\end{center}

\vskip 0.3in

\baselineskip 16pt
\date{}

\begin{center} {\bf Abstract} \end{center} 

We derive an extension of the Ryu-Takayanagi prescription for curvature squared theories of gravity in the bulk, and comment on a prescription for more general theories. This results in a new entangling functional, that contains a correction to Wald's entropy. The new term is quadratic in the extrinsic curvature. The coefficient of this correction is a second derivative of the lagrangian with respect to the Riemann tensor. For Gauss-Bonnet gravity, the new functional reduces to Jacobson-Myers'.
\vskip 0.2cm

\noindent 
\end{titlepage} \vfill\eject

\setcounter{equation}{0}

\pagestyle{empty}
\small
\normalsize
\pagestyle{plain}
\setcounter{page}{1}

\newpage

\section{Introduction}
Quantum entanglement appears to be a key ingredient for an understanding of holography \cite{VanRaamsdonk:2009ar, Swingle:2009bg}. The seminal proposal of Ryu and Takayanagi \cite{Ryu:2006bv, Ryu:2006ef} offers a way to reconstruct properties of a dual geometry from the spatial entanglement of a state. It conjectures that, for General Relativity in the bulk, the entanglement entropy $S$ of a region in a dual theory is the area $\mathcal{A}$ of a certain minimal surface in Planck units,
\beq
S=\frac{1}{4G}\mathcal{A}.
\label{RyuTakayangi}\eeq
In this paper we derive an analogous formula for curvature squared theories of gravity, and comment on another one that works more generally. Eventually, one would like to find a derivation for general theories of gravity, by which we mean theories of geometry:
\beq
I=\int \sqrt{g}\,d^D x\, \mathcal{L}(R_{\mu\nu\rho\sigma}, \nabla_\mu, g_{\mu\nu})+\textrm{Boundary terms}\,.
\label{GeneralAction}\eeq

We follow the approach of ref.~\cite{Lewkowycz:2013nqa}, that derived eq.~\eqref{RyuTakayangi} in General Relativity under the assumption that the euclidean field theory problem maps holographically to a geometric calculation in an euclidean manifold with $Z_n$ symmetry. The $n$ in the $Z_n$ is the number of replicas in the replica trick, that we will review below. In the classical gravity regime, the calculation of $S$ corresponds to a certain limit of the gravitational action of this manifold with $Z_n$ symmetry, $I_n$. One has to allow for a conical excess $2\pi\delta$ in the loci of fixed points of $Z_n$. In the action, we write the magnitude of this conical excess in brackets: $I_{n}[\delta]$. Then, the holographic calculation that one has to do is:
\beq
S=\left.\partial_n\left(I_{n}[0]-I_{n}[n-1]\right)\right|_{n=1}\,.
\label{LM}\eeq
Since the parameter $n$ was originally an integer (the number of replicas), the evaluation of this formula requires an analytic continuation of $I_n[n-1]$ to real $n$, that was discussed at some length in \cite{Lewkowycz:2013nqa}.

Things simplify in the presence of euclidean time translation symmetry. Euclidean stationarity corresponds to the enhancement of the $Z_n$ symmetry to $U(1)$. In this case, the derivation of $\eqref{RyuTakayangi}$ corresponds to a derivation of the entropy of Killing horizons in euclidean quantum gravity, that sees it as the quantity conjugate to a deficit angle \cite{Carlip:1993sa, Susskind:1994sm}. The entropy of such euclidean horizons in general theories of gravity is known to be given by Wald's formula \cite{Wald:1993nt, Iyer:1994ys, Jacobson:1993vj}
\beq
S=-2\pi\int_W \sqrt{\gamma}\,d^{D-2}\sigma\,\left.\epsilon_{\mu\nu}\epsilon_{\rho\sigma}\frac{\delta \mathcal{L}}{\delta R_{\mu\nu\rho\sigma}}\right|_{W}\,,
\label{Wald}\eeq
where $W$ is the loci of fixed points of the $U(1)$ symmetry and $\epsilon_{\mu\nu}$ is the binormal to $W$, which is a codimension$-2$ submanifold. In this formula the lagrangian has to be varied with respect to the Riemann tensor keeping the metric and connection fixed, as if they were independent quantities.

Eq.~\eqref{RyuTakayangi} is, however, valid beyond the requirement of euclidean time stationarity, and we would like to find a similarly robust formula for general theories \eqref{GeneralAction}. Stationarity in euclidean time is technically important for the derivation of \eqref{Wald}. The advantage is that it implies the vanishing of the extrinsic curvature of $W$. This can be seen directly by choosing coordinates adapted to $W$, and expanding in the distance to $W$:
\beq
ds^2=\left(\gamma_{ab}-2K_{ab}{}^1\, r\cos\theta-2K_{ab}{}^2\, r\sin\theta\right)d\sigma^a d\sigma^b+dr^2+r^2d\theta^2+\dots\,,
\label{EmbeddingLeading}\eeq
where ${\sigma^a}$, $a,b=1,\dots,D-2$, are coordinates in $W$, $\theta$ is the euclidean time, and $r$ is the distance to $W$. Clearly, euclidean time independence implies the vanishing of the extrinsic curvature $K_{ab}{}^i$.

We find that, for general embeddings of $W$ \eqref{EmbeddingLeading} and curvature-squared gravity\footnote{The case of more general gravitational lagrangians \eqref{GeneralAction} is discussed in the appendix following \cite{Dong:2013qoa}.}, formula \eqref{LM} gives\footnote{For other recent attempts see \emph{e.g.} \cite{ Chen:2013qma}, \cite{ Fursaev:2013fta}, \cite{Bhattacharyya:2013jma}.}:
\begin{equation}
S=\int_W \sqrt{\gamma}\,d^{D-2}\sigma\,\left.\left(\delta^{(1)}R_{\mu\nu\rho\sigma}\frac{\delta \mathcal{L}}{\delta R_{\mu\nu\rho\sigma}}+\delta^{(2)}R_{\mu\nu\rho\sigma\tau\pi\xi\zeta}\frac{\partial^2 \mathcal{L}}{\partial R_{\mu\nu\rho\sigma}\,\partial R_{\tau\pi\xi\zeta}}\right)\right|_{W}\,,
\label{NewEntropy}\end{equation}
where $\gamma_{ab}$ is the metric induced on $W$, and with
\begin{align}
\delta^{(1)}R_{\mu\nu}{}^{\rho\sigma}=&-4\pi\perp_{[\mu}{}^{\rho}\perp_{\nu]}{}^{\sigma}=-2\pi\,\epsilon_{\mu\nu}\epsilon^{\rho\sigma}\,,\label{delta1R}\\
\delta^{(2)} R_{\mu\nu}{}^{\rho\sigma}{}_{\tau\pi}{}^{\xi\zeta}=&4\pi
\left( K_{[\mu}{}^{[\rho|i|}\perp_{\nu]}{}^{\sigma]}{}_{[\pi}{}^{[\zeta}K_{\tau]}{}^{\xi]j}\perp_{ij}
+K_{[\mu}{}^{[\rho|k|}\tilde{\perp}_{\nu]}{}^{\sigma]}{}_{[\pi}{}^{[\zeta}K_{\tau]}{}^{\xi]l}\epsilon_{kl}\right)\,.\label{delta2R}
\end{align}
$\perp_{\mu}{}^{\nu}$ is the projector on the space transverse to $W$. We defined
\beq
\perp_{\nu\sigma\pi\zeta}=\perp_{\nu\pi}\perp_{\sigma\zeta}+\perp_{\nu\zeta}\perp_{\pi\sigma}-\perp_{\nu\sigma}\perp_{\pi\zeta}\,,\qquad 
\tilde{\perp}_{\nu\sigma\pi\zeta}=\perp_{\nu\pi}\epsilon_{\sigma\zeta}+\perp_{\sigma\zeta}\epsilon_{\nu\pi}\,.
\label{defperp4}\eeq

Note that the first term in the parenthesis of eq.~\eqref{NewEntropy} is Wald's entropy \eqref{Wald} and the second term vanishes in the absence of extrinsic curvature. Equations \eqref{NewEntropy}-\eqref{defperp4} are the main result of this paper, that is devoted to their derivation.

{\bf Note added in v2:} As the writing of this paper was being undertaken \cite{Dong:2013qoa} appeared in the arXiv, with a prescription that works for more general theories of gravity. We discuss that prescription in the appendix.

\section{Entanglement entropy and its holographic dual}
This section reviews some aspects of the connection between entanglement and geometry. This includes a derivation of eq.~\eqref{LM} and relevant arguments in favor of the Ryu-Takayangi conjecture that were developed in \cite{Lewkowycz:2013nqa}.

The entanglement entropy of a density matrix\footnote{$\rho$ could be pure: $\rho=|\psi\rangle\langle\psi|$.}  $\rho$ in a spatial region $A$ is computed as the Von Neumann entropy of its reduced density matrix $\rho_A$. This, in turn, is obtained from tracing over the degrees of freedom outside $A$: $\rho_A=\tr_{\bar{A}}\rho$. Then:
\beq
S=-\tr \left(\rho_A\log\rho_A\right)\,.
\label{SVN}\eeq

It is difficult to compute $S$ directly. A standard tool to calculate it indirectly is the replica trick, which connects $S$ to geometry already in the field theory picture \cite{Callan:1994py}. One needs to first consider the R\'{e}nyi entropies:
\beq
S_{n}=\frac{-1}{n-1}\log\left(\tr\rho_A^n\right)\,,
\eeq
which are defined for $n\in \mathbb{Z}$, and analytically continue them to $n\in \mathbb{R}$. Then, one exploits that
\beq
S=\lim_{n\rightarrow 1}S_n
\eeq
to calculate $S$.

There are ambiguities when continuing functions from $\mathbb{Z}$ to $\mathbb{R}$, as one can add terms of the type $\sin\pi n$, which vanish for all $n\in \mathbb{Z}$ but not for real $n$. A prescription is needed to quotient out these ambiguities (usually, regularity as $n\rightarrow \pm i \infty$).

A geometric interpretation of the R\'{e}nyi entropies $S_n$ (and therefore $S$) goes as follows: If the state $\rho$ is generated via an euclidean path integral on the space $A\cup \bar{A}$ times euclidean time $\theta$, then $\rho_A^n$ can be generated via that path integral on $(A\otimes n\, \theta)\cup (\bar{A}\otimes \theta)$. By this notation it is meant that for each extra power of $\rho_A$ in $S_n$ one evolves in $A$ for an extra interval of $\theta$  relative to $\bar{A}$ and glues consecutive evolutions together, hence the $Z_n$ symmetry. Finally, to take the trace one needs to identify the endpoints of the time interval, thus closing euclidean time in a loop. This results in a path integral in an euclidean manifold with a compact euclidean time circle that has a period in $A$ that is $n$ times that in $\bar{A}$. Continuity along $\partial A$ requires the euclidean time circle to close off at $\partial A$, thus creating a conical excess on $\partial A$, of opening angle $2\pi (n-1)$.

In conformal field theories $\partial A$ can be pushed to infinity by a conformal transformation for simple enough regions \cite{Casini:2011kv, Lewkowycz:2013nqa}. In the following we will assume this has been done, so there will be no conical singularities in the field theory side.

Holography maps this euclidean field theory problem to an euclidean gravity one in one more dimension. In the usual limit of classical gravity in the bulk, one has
\beq
\tr\rho_A^n\approx\frac{e^{-I_n}}{e^{-nI_1}}\,,
\label{dualrhon}\eeq
where $I_n$ indicates the value of the gravitational action on the solution of the equations of motion with an euclidean time with period $2\pi n$. $I_1$ is the solution with period $2\pi$ and computes $\tr\rho$, which we use to normalize the gravitational calculation. We assume we work with states with a good holographic dual, such that the geometries in $I_n$ and $I_1$ are both everywhere regular. The entanglement entropy $S$ \eqref{SVN} is then related to the limit of the analytical continuation of $I_n$ for $n\in \mathbb{R}$ as $n\rightarrow 1$:
\beq
S=\left.\partial_n\left(I_n-n I_1\right)\right|_{n=1}\,.
\eeq

Using locality of the action the factor of $n$ in the second term in the rhs can be absorbed in the period of the euclidean time. This allows to rewrite this term as the action of a manifold with euclidean time period $2\pi n$, with a conical excess of $2\pi(n-1)$ where the time circle shrinks to zero size\footnote{We called this submanifold $W$ in the introduction and will continue to do so in the rest of the text.}, without including any contribution from the conical singularity \cite{Lewkowycz:2013nqa}. We call this quantity $I_{n}[n-1]$; the subindex refers to the period of the euclidean time and the argument in brackets refers to the conical excess on $W$. Then eq.~\eqref{LM} follows
\beq
S=\left.\partial_n\left(I_n[0]- I_n[n-1]\right)\right|_{n=1}\,.
\label{HoloS}\eeq
One can rewrite $I_n[n-1]$ as
\beq
I_n[n-1]={}^{(F)}I_n[n-1]-\hat{I}_n[n-1]\,,
\eeq
where ${}^{(F)}I_n[n-1]$ is the full action of the conically singular manifold and $\hat{I}_n[n-1]$ is the contribution to the action from the conical singularity. It will be convenient to think about these conical manifolds as the limit of families of regular geometries. Each of these geometries is regular and its action differs from $I_n[0]$ by terms of order $(n-1)^2$, as the geometry in $I_n[0]$ satisfies the equations of motion. Then, $I_n[0]$ and ${}^{(F)}I_n[n-1]$ cancel in the $n\rightarrow 1$ limit and eq.~\eqref{HoloS} reduces to
\beq
S=\hat{I}_1^\prime[0]\,,
\label{SCone}\eeq
where the prime derives the argument in brackets, not the subscript.

Eq.~\eqref{SCone} instructs us to isolate the contributions to the action that are linear in the conical excess $(n-1)$ and non-extensive, that is, that do not come from the integral in $\theta$ running from $0\leq \theta<2\pi n$ instead of $0\leq\theta<2\pi$. These contributions will naturally localize on $W$, and this will give the holographic entanglement entropy functional. This calculation is done in the next section. 
However, we should emphasize that the conical geometry appears only as a tool to compute $n I_1[0]$. The holographic calculation of the R\'{e}nyi entropies can be phrased solely in terms of regular holographic duals \eqref{dualrhon} \cite{Headrick:2010zt, Hung:2011nu}.

\section{Action of conical singularities}
In this section we develop a framework to evaluate eq.~\eqref{SCone} for general theories of gravity \eqref{GeneralAction} and general embeddings of $W$ \eqref{EmbeddingLeading}. As a warm up, we start with a review of the case without extrinsic curvature.

\subsection{Zero extrinsic curvature case}
Because of their localized nature, conical singularities are subtle to deal with using differential geometric tools. However, it is known that if they do not possess extrinsic curvature they can be accounted for as delta-like contributions in the Riemann tensor, to leading order in the conical excess  \cite{Fursaev:1995ef}. For an excess of $2\pi(n-1)$ this contribution is
\beq
\delta R_{\mu\nu\rho\sigma}=-2\pi(n-1)\epsilon_{\mu\nu}\epsilon_{\rho\sigma}\,\delta^{(2)}(W)\,,
\label{deltaRnoK}\eeq
where $\epsilon_{\mu\nu}$ is the binormal to $W$, on which the Dirac delta localizes. Wald's entropy \eqref{Wald} follows straightforwardly from the application of \eqref{deltaRnoK} to eq.~\eqref{SCone}, \cite{Myers:2010tj}.

One can derive eq.~\eqref{deltaRnoK} from the following construction. Consider the metric around a point in $W$, in the absence of extrinsic curvature:
\beq
ds^2=dr^2+r^2d\theta^2+\delta_{ab}d\sigma^a d\sigma^b+O(1/\lambda^2)\,,
\label{EmbeddingNoK}\eeq
where we have taken normal coordinates $\sigma^a$ in $W$ around the chosen point. $\lambda$ is the curvature lengthscale of $W$ or the background, of which some components are in fact related by Gauss-Codacci equations. The vanishing of the extrinsic curvature term, that we allowed for in \eqref{EmbeddingLeading}, corresponds to the absence of $1/\lambda$ terms in \eqref{EmbeddingNoK}. This guarantees that normal coordinates in $W$ are also part of a normal chart of the background. Indeed, the extrinsic curvature would measure the failure of geodesics of $W$ being geodesics of the background and therefore of normal coordinates grids in $W$ being normal coordinate grids in the background. $K_{ab}{}^i=0$ is what makes eq.~\eqref{EmbeddingNoK} look like flat space in Cartesian coordinates, something that will not be true in the next section nor in the appendix (despite the  fact that we will choose again normal coordinates in $W$ and its transverse space).

We want to introduce a conical excess on $W$ in \eqref{EmbeddingNoK} by periodically identifying the euclidean time $\theta$ with period $2\pi n$ instead of $2\pi$. A convenient way to do so is considering a family of geometries regulating the conical singularity on scales $r\lesssim\Lambda\ll\lambda$:
\beq
ds^2=dr^2+r^2\left(1-\frac{n-1}{n}A(r^2/\Lambda^2)\right)^2d\theta^2+\delta_{ab}d\sigma^a d\sigma^b+O(1/\lambda^2)\,,
\label{EmbeddingNoKRegular}\eeq
where $A(r^2/\Lambda^2)$ is a regulating function, decaying sufficiently fast for $r\gg\Lambda$ and such that $A(0)=1$. $A(r^ 2/\Lambda^2)$ is an off-shell deformation of the geometry \eqref{EmbeddingNoK} on a scale of order $\Lambda$, that restores regularity around $r=0$ when $\theta\sim\theta+2\pi n$ . Eq.~\eqref{EmbeddingNoKRegular} is, thus, a regulated version of the conical excess in \eqref{EmbeddingNoK}. Our strategy will be to compute on this geometry and send the cutoff $\Lambda$ to $0$ at the end.

To evaluate eq.~\eqref{SCone} one needs to isolate the explicit dependence on $n$ coming from the integrand in $I$ (and not from the limits of integration $0\leq \theta < 2\pi n$). Because this dependence appears in \eqref{EmbeddingNoKRegular} multiplied by the localizing function $A(r^2/\Lambda^2)$, the integral we are interested in will localize on scales $r\lesssim \Lambda$. Since at the end of the day we will take the limit $\Lambda\rightarrow 0$, all $\Lambda$ dependence should be made explicit. This can be done by going to coordinates $r=\Lambda y$. The measure in \eqref{GeneralAction} reads:
\beq
\sqrt{g}\,d^D x=\left(1-\frac{n-1}{n}A(y^2)\right)\Lambda^2\,y\, d\theta\, dy\,d^{D-2}\sigma  +O(\Lambda^3)\,.
\label{measure}\eeq

In the limit $\Lambda\rightarrow 0$ we will get a finite $n-$dependent contribution to $I$ from the integrand if this has a $O(1/\Lambda^2)$ contribution. This is best assessed in an orthonormal frame, where powers of $\Lambda$ coming from $g_{\theta\theta}$ and $g^{\theta\theta}$ are automatically accounted for. We thus pick the basis
\beq
e^{\hat{r}}=dr\,,\qquad e^{\hat{\theta}}=r\left(1-\frac{n-1}{n}A(r^2/\Lambda^2)\right)d\theta\,,\qquad e^{\hat{\sigma}^a}=d\sigma^a
\label{FrameNoK}\eeq
and compute the Riemann tensor. The only $1/\Lambda^2$ contribution is
\beq
R_{\hat{r}\hat{\theta}\hat{r}\hat{\theta}}=\frac{1}{\Lambda^2}\frac{n-1}{n}\frac{1}{y\left(1-\frac{n-1}{n}A(y^2)\right)}\frac{d^2(y A(y^2))}{dy^2}\,,
\eeq
and identically in other components related to this one by reordering of the indices using symmetries.

At this stage it is useful to recall that to compute the entropy from \eqref{SCone}, we are only interested in contributions linear in $(n-1)$ to the action \eqref{GeneralAction}. To that effect, $R_{\hat{r}\hat{\theta}\hat{r}\hat{\theta}}$ can contribute at most only linearly\footnote{Crucially, this kind of argument will fail in the next subsection, but it is correct here.}. Then, at small $\Lambda$
\beq
\hat{I}=\sum_{\textrm{ordering}}\int d\theta\, dy\,d^{D-2}\sigma\,(n-1)\frac{d^2(y A(y^2))}{dy^2}\frac{\delta \mathcal{L}}{\delta{R_{\hat{r}\hat{\theta}\hat{r}\hat{\theta}}}}+O(\Lambda)+O\left((n-1)^2\right)\,,
\eeq
where the sum is for all the ordering of the $\hat{r}\hat{\theta}\hat{r}\hat{\theta}$ indices, and corresponding sign weights. As advertised, the integral over $y$ localizes on the support of $A(y^2)$, on which the derivative of the lagrangian is effectively constant. This integral can be done with independence of the choice of regulating function $A(y^2)$, provided it satisfies the conditions we stated below eq.~\eqref{EmbeddingNoKRegular}. Taking the $\Lambda\rightarrow 0$ limit and performing the integrals in $\theta$ and $y$, and covariantazing
\beq
\hat{I}=\int_W\sqrt{\gamma}\,d^{D-2}\sigma\,\left(-2\pi(n-1)\epsilon_{\mu\nu}\epsilon_{\rho\sigma}\right)\left.\frac{\delta \mathcal{L}}{\delta{R_{\mu\nu\rho\sigma}}}\right|_{W}+O\left((n-1)^2\right)\,,
\eeq
from which eq.~\eqref{deltaRnoK} follows.

Note that the considerations in this subsection are independent of $I$ being the action in \eqref{SCone}, and apply to any local functional of the curvature. 

\subsection{General case}\label{sec:BentCone}
We now address the general case \eqref{EmbeddingLeading}, allowing for non-zero extrinsic curvature. We will refer to this setup as a `bent cone'.\footnote{In the literature it is sometimes also referred to as `squashed cone' \cite{Dowker:1994bj}.} The strategy we follow is exactly analogous to the one in the previous section. That is, we seek for a regulator of the bent cone at some distances $r\lesssim \Lambda\ll\lambda$.

A good regulating geometry is given by\footnote{This is motivated by good properties of the geometries introduced in \cite{Fursaev:2013fta}, which were further studied in \cite{Bhattacharyya:2013gra}. See also \cite{Alishahiha:2013zta}.}
\beq
\begin{split}
ds^2=\left(\delta_{ab}-\frac{2}{\lambda}\left(K_{ab}{}^1\, r\cos\theta+K_{ab}{}^2\, r\sin\theta\right)\left(\frac{r}{\Lambda}\right)^{(n-1)B(r^2/\Lambda^2)}\right)&d\sigma^a d\sigma^b\\
+dr^2+r^2\left(1-\frac{n-1}{n}A(r^2/\Lambda^2)\right)^2 d\theta^2+&O(1/\lambda^2)\,,
\end{split}
\label{GeneralRegulated}\eeq
with $B(y^2)$ another regulating function with the same defining properties as $A(y^2)$. We have again kept $\lambda$ as a book-keeping parameter of the curvature lengthscale.

The $r-\theta$ part of the metric in eq.~\eqref{GeneralRegulated} is regulated exactly in the same way as in eq.~\eqref{EmbeddingNoKRegular}. The difference lays in the extrinsic curvature terms, that we now allowed. To justify the regulation by $B(y^2)$, consider equation \eqref{GeneralRegulated} in an expansion in $r/\Lambda$:
\beq
ds^2=\left(\delta_{ab}-\frac{2}{\lambda}\left(\frac{r}{\Lambda}\right)^{n-1}\left(K_{ab}{}^1\, r\cos\theta+K_{ab}{}^2\, r\sin\theta\right) 
\right)d\sigma^a d\sigma^b
+dr^2+\frac{r^2}{n^2} d\theta^2+O(r^2/\Lambda^2)\,,
\label{GeneralRegulatedExp}
\eeq
and go to Cartesian coordinates in the directions transverse to $W$:
\beq
x^1=r\cos\left(\frac{\theta}{n}\right)\,,\qquad x^2=r\sin\left(\frac{\theta}{n}\right)\,,
\label{ConicalCartesianCoords}\eeq
where we used that $\theta/n$ is the true polar coordinate in \eqref{GeneralRegulated}, with period $2\pi$.\footnote{This makes explicit the $Z_n$ symmetry in \eqref{GeneralRegulated}, $\theta\rightarrow\theta+2\pi k$.} The metric looks:
\beq
ds^2=\left(\delta_{ab}-\frac{2}{\lambda}\frac{r^n}{\Lambda^{n-1}}\left(K_{ab}{}^1\, T_n (x^1/r)+K_{ab}{}^2\, \tilde{T}_n(x^2/r)\right) 
\right)d\sigma^a d\sigma^b
+(dx^1)^2+(dx^2)^2+O(r^2/\Lambda^2)\,,
\label{GeneralRegulatedx1x2}
\eeq
with $T_n$ the Chebyshev polynomials, defined by
\beq
T_n(\cos\theta)=\cos(n\theta)\,,
\eeq
and $\tilde{T}_n$ defined similarly for the sines. The $T_n$ have degree and parity $n$, so the metric \eqref{GeneralRegulatedx1x2} is explicitly regular around the origin for integer $n$.\footnote{\emph{e.g.} for $n=2$, this metric \eqref{GeneralRegulatedx1x2} is
\beq
ds^2=\left(\delta_{ab}-\frac{2}{\lambda}\frac{1}{\Lambda}\left(K_{ab}{}^1\, \left((x^1)^2-(x^2)^2\right)+K_{ab}{}^2\, 2\,x^1 x^2\right) 
\right)d\sigma^a d\sigma^b
+(dx^1)^2+(dx^2)^2+O(r^2/\Lambda^2)\,.\nonumber
\eeq
} It is crucial to observe that for this to happen the power of $r$ in front of the extrinsic curvature terms in eq.~Ê\eqref{GeneralRegulatedExp} has to be $n$, thus justifying the presence of the $B(y^2)$ terms in \eqref{GeneralRegulated}. Otherwise, we get overall fractional powers of $r$ for integer $n$ in \eqref{GeneralRegulatedx1x2}, explicitly spoiling regularity at $r=0$.  An analogous analysis holds for the $\tilde{T}_n$.

Notice that choosing $B(0)$ to be any odd positive integer would also lead to regular geometries for integer $n$. Therefore, there seems to be a degree ambiguity in the regulation. In the remainder of the main text we will work in the minimal case $B(0)=1$. See sec.~\ref{sec:comments} for comments about this.

As in the previous subsection, we now wish to find $1/\Lambda^2$ contributions to the integrand \eqref{GeneralAction}. On the one hand, there will be the same contribution coming from the $r-\theta$ part of the metric \eqref{GeneralRegulatedExp} that we discussed in the previous subsection, namely Wald's entropy. This is the origin of the first term in \eqref{NewEntropy}. On the other hand, we expect new contributions, coming from having allowed $W$ to have extrinsic curvature. Consider, then, the possible scalings with $1/\Lambda$ of these terms. Because the extrinsic curvature has dimensions of one over length and the Riemann is one over length squared, dimensional analysis says that any such contribution to the Riemann tensor will be, at best, of the type
\beq
\textrm{Riem}\sim(n-1)\frac{K}{\Lambda}\,,
\label{ScalingRiemannK}\eeq
where we have made explicit the leading dependence in $n-1$: A term like \eqref{ScalingRiemannK}, with a factor of $1/\Lambda$, will be at least proportional to $(n-1)$, because it comes from having a (regulated) conical excess that vanishes as $n\rightarrow 1$.

Because of the arguments around eq.~\eqref{measure}, in the $\Lambda\rightarrow 0$ limit only terms scaling as $1/\Lambda^2$ in the integrand will give a finite result, so terms like \eqref{ScalingRiemannK} can contribute to the integral \eqref{GeneralAction} if they appear squared. This is the origin of the second derivative of the lagrangian in \eqref{NewEntropy}.

One may now be puzzled by the following fact. We are ultimately interested in eq.~\eqref{LM}, which instructs us to take one derivative of the action with respect to $(n-1)$, and send $n\rightarrow 1$. It thus seems impossible that a product of two terms \eqref{ScalingRiemannK}, which will be at least $O((n-1)^2)$, gives a finite result in the required limit. There is, however, one subtlety, coming from extra $n-$dependence in the integrand. This is best illustrated with an explicit example.

Consider the following calculation
\beq
\lim_{n\rightarrow 1}\partial_n\int_0^{\infty}dy\, (n-1)^2y^{2n-3}A(y^2)\,,
\label{Examplento1}\eeq
where $A(y^2)$ is one of the regulating functions we have discussed. If one commutes the order of the integral and the limit, the result is naively zero. However, this integral diverges in the lower end as $n\rightarrow 1$. We need to perform the integral for $n>1$, analytically continue the result to $n\sim 1$, and then take the derivative and the limit $n\rightarrow 1$.

The result of this procedure is independent of the choice of the regulating function $A(y^2)$, so let us choose one for which the calculation is straightforward. For $A(y^2)=e^{-y^2}$, the integral can be done in terms of Gamma functions, resulting in
\beq
\lim_{n\rightarrow 1}\partial_n\int_0^{\infty}dy\, (n-1)^2y^{2n-3}e^{-y^2}=\lim_{n\rightarrow 1}\partial_n\left((n-1)^2\frac{\Gamma(n-1)}{2}\right)=\frac{1}{2}\,.
\eeq
This is exactly the mechanism by which terms like squares of \eqref{ScalingRiemannK} will contribute in \eqref{LM}.

Let us now do the calculation we set to do. Again, we calculate in the orthonormal frame of \eqref{FrameNoK}, with the change
\beq
e^{\hat{\sigma}^a}=d\sigma^b\left(\delta^{a}{}_b-\frac{1}{\lambda}\left(K^a{}_b{}^1\, r\cos\theta+K^a{}_b{}^2\, r\sin\theta\right) \left(\frac{r}{\Lambda}\right)^{(n-1) B(r^2/\Lambda^2)}\right)+O(1/\lambda^2)\,.
\eeq
The contributions to the Riemann of the interesting type \eqref{ScalingRiemannK} are
\begin{align}
\delta R_{\hat{\sigma}^a\hat{r}\hat{\sigma}^b\hat{r}}=&\frac{n-1}{\lambda\, \Lambda}\left(K_{ab}{}^1\cos\theta+K_{ab}{}^2\sin\theta\right)y^{(n-1)B(y^2)-1}\nonumber\\
&\times\left[B(y^2)+2y^2(2+3\log y)B^\prime(y^2)+4y^4\log y\,B^{\prime\prime}(y^2))+O(n-1)\right]\,,
\label{Rarbr}
\\
\delta R_{\hat{\sigma}^a\hat{\theta}\hat{\sigma}^b\hat{r}}=&\frac{n-1}{\lambda\, \Lambda}\left(-K_{ab}{}^1\sin\theta+K_{ab}{}^2\cos\theta\right)y^{(n-1)B(y^2)-1}\nonumber\\
&\times\left[B(y^2)+2y^2(A^\prime(y^2)+\log y\, B^\prime(y^2))+O(n-1)\right] \,,
\label{Rathetabr}
\end{align}\begin{align}
\delta R_{\hat{\sigma}^a\hat{\theta}\hat{\sigma}^b\hat{\theta}}=
&-\frac{n-1}{\lambda\, \Lambda}\left(K_{ab}{}^1\cos\theta+K_{ab}{}^2\sin\theta\right)y^{(n-1)B(y^2)-1}\nonumber\\
&\times \left[2A(y^2)-B(y^2)+2y^2(A^\prime(y^2)-\log y\, B^\prime(y^2))+O(n-1)\right]\,,
\label{Rathetabtheta}\end{align}
and other components related to these by reordering of indices by symmetries.

A word of caution is due at this point. Here we are not expanding around the $n=1$ case, as in that expansion the integrals we will do are divergent, as explained around eq.~\eqref{Examplento1}. We are just writing conveniently the $n$ dependence of these terms. The setup to have in mind is performing the integrals for integer $n\ge 2$, where they converge and the geometry is well defined, and then analytically continue the result to $n\sim 1$. After doing this, only the terms coming from the ones sown explicitly above will contribute to the limit $n\rightarrow 1$.

In the remainder of this section we will neglect possible further $y$ dependence coming from the second derivative of the lagrangian. This assumption limits the validity of the derivation to curvature squared gravity, where $\partial^2\mathcal{L}/\partial\textrm{Riem}^2$ is a function of the metric only, and lets us factor out this term outside the $\theta$ and $y$ integrals. Considering possible $y$ dependence coming from it enlarges the domain of applicability of the derivation (\emph{cf.} \cite{Dong:2013qoa}, see also appendix).

The integrands we are interested in are then squares of \eqref{Rarbr}-\eqref{Rathetabtheta}. We will refer to them by their $\hat{r}-\hat{\theta}$ indices, as the $\hat{\sigma}^a-\hat{\sigma}^b$ ones factor out trivially in extrinsic curvatures. The integral on $\theta$ is effectively done with the prescription in \cite{Lewkowycz:2013nqa}. We first integrate from $0\le\theta<2\pi$, and then multiply the result by $n$. This averages some pairings of the extrinsic curvature terms in \eqref{Rarbr}-\eqref{Rathetabtheta} to zero, letting survive two types: those that have their third index effectively contracted, $K_{ab}{}^i K_{cdi}$, and those that are effectively antisymmetrized, $K_{ab}{}^i K_{cd}{}^{j}\epsilon_{ij}$.\footnote{Our choice of transverse orientation is $\epsilon_{x^1 x^2}=+1$. This corresponds to $\epsilon_{\hat{r}\hat{\theta}}=+1$.} Of the ones that effectively have the third index contracted, the only ones giving non-zero are:
\beq
\hat{r}\hat{r}\,\hat{r}\hat{r}\,,\qquad \hat{r}\hat{r}\,\hat{\theta}\hat{\theta}\,,\qquad \hat{r}\hat{\theta}\,\hat{r}\hat{\theta}\,,\qquad \hat{\theta}\hat{\theta}\,\hat{\theta}\hat{\theta}\,.
\label{ListIntegrals}\eeq
Of the antisymmetrized combinations, the surviving ones are
\beq
\hat{r}\hat{r}\,\hat{r}\hat{\theta}\,,\qquad \hat{\theta}\hat{\theta}\,\hat{r}\hat{\theta}\,.
\eeq

The integral on $y$ is independent of the regulating functions and can be done by noticing that the contribution that will matter when we take the limit $n\rightarrow 1$ comes entirely from the terms at the origin of the integrals, $y=0$. To that effect one can keep only the terms not explicitly multiplied by $y$ in the second lines of \eqref{Rarbr}-\eqref{Rathetabtheta}. Also for that matter, the exponents of $y$ are effectively their value at $y=0$, so the $B(y^2)$ in the exponents are effectively $1$. Then, the integrals \eqref{ListIntegrals} become analogous to the one we discussed explicitly in \eqref{Examplento1}, with possibly the $A$s exchanged by $B$s. After the dust settles, the result of taking the derivative in $n$ of the integral, and sending $n\rightarrow 1$ is:
\beq
\hat{r}\hat{r}\,\hat{r}\hat{r}=\frac{\pi}{2}\,,\qquad \hat{r}\hat{r}\,\hat{\theta}\hat{\theta}=-\frac{\pi}{2}\,,\qquad \hat{r}\hat{\theta}\,\hat{r}\hat{\theta}=\frac{\pi}{2}\,,\qquad \hat{\theta}\hat{\theta}\,\hat{\theta}\hat{\theta}=\frac{\pi}{2}\,,
\qquad \hat{r}\hat{r}\,\hat{r}\hat{\theta}=\frac{\pi}{2}\,,\qquad \hat{\theta}\hat{\theta}\,\hat{r}\hat{\theta}=-\frac{\pi}{2}\,.
\eeq
where the above notation refers to, \emph{e.g.},
\beq
\lim_{n\rightarrow 1}\partial_n\int\sqrt{g}\,d^Dx\,\delta R_{\hat{\sigma}^a\hat{r}\hat{\sigma}^b\hat{r}}
\delta R_{\hat{\sigma}^c\hat{\theta}\hat{\sigma}^d\hat{\theta}}
=\hat{r}\hat{r}\,\hat{\theta}\hat{\theta}\,\int_W\sqrt{\gamma}\,d^{D-2}\sigma K_{ab}{}^i K_{cd i}\,,
\label{covintcont}\eeq
and
\beq
\lim_{n\rightarrow 1}\partial_n\int\sqrt{g}\,d^Dx\,\delta R_{\hat{\sigma}^a\hat{r}\hat{\sigma}^b\hat{r}}
\delta R_{\hat{\sigma}^c\hat{r}\hat{\sigma}^d\hat{\theta}}
=\hat{r}\hat{r}\,\hat{r}\hat{\theta}\,\int_W\sqrt{\gamma}\,d^{D-2}\sigma K_{ab}{}^i K_{cd}{} ^{j}\epsilon_{ij}\,,
\label{covintantisym}\eeq
and likewise for the other combinations.

This final result \eqref{covintcont}-\eqref{covintantisym} can be summarized covariantly as
\beq
\lim_{n\rightarrow 1}\partial_n\int_0^{2\pi n}d\theta\int_0^\infty dy\,y\, \delta R_{a\nu b\sigma}\delta R_{c\pi d\zeta}=\frac{\pi}{2}K_{ab}{}^iK_{cd}{}^{j}\left(\perp_{\nu\sigma\pi\zeta}\perp_{ij}+\tilde{\perp}_{\nu\sigma\pi\zeta}\epsilon_{ij}\right)\,,
\label{covmS}\eeq
where we brought back the factors of the extrinsic curvature, and defined
\beq
\perp_{\nu\sigma\pi\zeta}=\perp_{\nu\pi}\perp_{\sigma\zeta}+\perp_{\nu\zeta}\perp_{\pi\sigma}-\perp_{\nu\sigma}\perp_{\pi\zeta}\,,\qquad 
\tilde{\perp}_{\nu\sigma\pi\zeta}=\perp_{\nu\pi}\epsilon_{\sigma\zeta}+\perp_{\sigma\zeta}\epsilon_{\nu\pi}\,.
\eeq
$\perp_{\mu\nu}$ is the projector in the space transverse to $W$, and $\epsilon_{\mu\nu}$ its binormal.

The result in \eqref{NewEntropy}  follows immediately from a symmetrization of the indices in \eqref{covmS}.

\section{Curvature squared gravity}
In this section we benchmark the prescription \eqref{NewEntropy} against Einstein-Gauss-Bonnet. This theory has as bulk lagrangian
\beq
I_{GB}=-\frac{1}{16\pi G}\int \sqrt{g}\,d^D x\, R-\lambda\int \sqrt{g}\,d^D x\,\left(R^2-4 R_{\mu\nu}R^{\mu\nu}+R_{\mu\nu\rho\sigma}R^{\mu\nu\rho\sigma}\right)\,.
\eeq

It is straightforward to compute what the entropy \eqref{NewEntropy} is for each of these terms. We get
\begin{align}
S_{\textrm{Riem}^2}&=-8\pi\int_W \sqrt{\gamma}\,d^{D-2} x\, \left(\perp_{\mu\rho}\perp_{\nu\sigma}R^{\mu\nu\rho\sigma}-K_{ab}{}^iK^{ab}{}_{i}\right)\,,
\label{Riemsq}\\
S_{\textrm{Ricci}^2}&=-4\pi\int_W \sqrt{\gamma}\,d^{D-2} x\, \left(\perp_{\mu\rho}R^{\mu\rho}-\frac{1}{2}K^iK_{i}\right)\,,
\label{Ricsq}\\
S_{\textrm{R}^2}&=-8\pi\int_W \sqrt{\gamma}\,d^{D-2} x\, R\,,
\label{Rsq}\end{align}
in agreement with \cite{Fursaev:2013fta}. We defined $K^i=\gamma^{ab}K_{ab}{}^i$.

Using a Gauss-Codacci relation, \eqref{Riemsq}, \eqref{Ricsq}, \eqref{Rsq} combine into the entropy
\beq
S_{GB}=S_{JM}=\frac{\mathcal{A}}{4G}+8\pi\lambda\int_W \sqrt{\gamma}\,d^{D-2} \sigma\,{}^{(W)}R\,,
\label{EntropyGB}\eeq
with $\mathcal{A}$ the area of $W$ and $^{(W)}R$ the Ricci scalar of $\gamma_{ab}$, the metric induced on $W$. This is the well known Jacobson-Myers functional \cite{Jacobson:1993xs}, that has been argued to have good features as a holographic entanglement entropy functional in these theories \cite{Hung:2011xb, deBoer:2011wk}, and whose lorentzian version is a candidate to horizon entropy in Lovelock theories for non-stationary situations \cite{Jacobson:1993xs}.

\section{Comments}\label{sec:comments}
We have derived a formula \eqref{NewEntropy} for the generalized gravitational entropy of \cite{Lewkowycz:2013nqa} for curvature squared theories of gravity from first principles. This is a functional that localizes on $W$, which is the loci of fixed points of the $Z_n$ symmetry in the geometries dual to the replica trick. In practice, $W$ can be found by minimizing the entropy functional \eqref{NewEntropy}, as the action around $n\sim 1$ is essentially the coupling of \eqref{GeneralAction} and $(n-1)$ times \eqref{NewEntropy}, and the replica geometries must satisfy the gravitational equations of motion, which extremize that action.\footnote{This argument assumes implicitly that the relevant conically singular geometries can be produced by sourcing the gravitational equations with the stress tensor of \eqref{NewEntropy}. This is true for $R^k$ gravity and for Lovelock densities, where this coupling in the equations of motion produces the desired on-shell cancellation of contributions localised on $W$, but we do not have a general argument. For a more thorough analysis of the equations of motion see \cite{Dong:2013qoa}.}

For more general theories of gravity, \eqref{NewEntropy} needs a modification \cite{Dong:2013qoa} that we discuss in the appendix.

Let us go back to the issue of what fixes the value $B(0)=1$, mentioned in sec.~\ref{sec:BentCone}. We remind the reader that demanding regularity at the origin for integer $n$ only fixes $B(0)$ to be a positive odd integer. A different value would change the result of the integrals in \eqref{ListIntegrals}, because it would change the factor multiplying $n$ in the exponent in eq.~\eqref{Examplento1}. Therefore, the final prescription \eqref{NewEntropy} would be different. We have chosen to fix this with the minimal prescription $B(0)=1$. In the case of Einstein gravity, this value for $B(0)$ could be chosen by demanding that for $n\sim1$ the gravitational equations of motion have no singularity as $r\rightarrow 0$. More specifically, demanding no $1/r$ divergence at $O(n-1)$ in the $a-b$ components of the Ricci tensor following from eqs.~\eqref{Rarbr}-\eqref{Rathetabtheta} sets $B(0)=A(0)=1$.\footnote{I thank the referee for pointing this out.} This argument, however, does not apply more generally, and we choose instead the minimal prescription.

As in \cite{Lewkowycz:2013nqa}, we have assumed that the replica symmetry $Z_n$ is not broken. It would be very interesting to drop this assumption. 

Since we worked in the frame where the $\partial A$ (the boundary of the entangling region in the field theory) had been pushed to infinity, we missed the global part of the Ryu-Takayangi prescription (and its generalization \eqref{NewEntropy}), that demands that the holographic entangling surface is homologous to $\partial A$. It would be very interesting to extend the setup to recover this.

We would also like to use the prescription \eqref{NewEntropy} to learn about the structure of the universal (logarithmic) divergences in the entanglement entropy of $6D$ CFTs through the AdS/CFT correspondence.\footnote{The corresponding problem in $4D$ CFTs was studied in \cite{Solodukhin:2008dh}.} We plan to address this in the future.

Reference \cite{Armas:2013hsa} suggests that the coefficients of terms quadratic in the extrinsic curvature in effective actions for extended objects can be interpreted as elastic moduli. It would be interesting to see if this interpretation can be pursued in \eqref{NewEntropy} to characterize physical properties of entangling surfaces.

It would also be interesting to study the implication of our results for the proposal of \cite{Bianchi:2012ev, Myers:2013lva}.

Finally, it is a pressing question to have a derivation of these questions in a purely lorentzian setup. There, prescriptions for the holographic dual of the entanglement entropy exist \cite{Hubeny:2007xt}, and there is a conjecture for a good notion of Noether entropy in time dependent setups in any theory of gravity \cite{Iyer:1994ys}. The lorentzian version of \eqref{NewEntropy} is different from this Iyer-Wald formula, as can be seen, \emph{e.g.}, by inspecting the case of $f(R)$ gravity. In this case \eqref{NewEntropy} predicts the entropy conjectured in \cite{oai:arXiv.org:gr-qc/9503020}, that is different from the one following from the Iyer-Wald construction, as noticed in \cite{oai:arXiv.org:gr-qc/9503020}. In that context, it would be interesting to study more generally in what circumstances the new entropy satisfies a second law.

\section*{Acknowledgements}
We gratefully acknowledge Don Marolf, Rob Myers, Eric Perlmutter and Misha Smolkin for discussions; Mariano Chernicoff, Don Marolf and Rob Myers for comments on a draft; Xi Dong for correspondence; and an anonymous referee for constructive criticisms. Work supported by the European Research Council grant no. ERC-2011-StG 279363-HiDGR.

\appendix
\section{The prescription in \cite{Dong:2013qoa}}
This section reviews the results in \cite{Dong:2013qoa} using the notation of the main text. That paper does not assume the $y$ independence of $\partial^2\mathcal{L}/\partial\textrm{Riem}^2$ that we assumed below eq.~\eqref{Rathetabtheta}, and therefore finds a more generally applicable entropy formula. The domain of validity of that derivation includes general theories without explicit derivatives of the curvature in the Lagrangian, $\mathcal{L}(g_{\mu\nu}, R_{\mu\nu\rho\sigma})$.

The paper \cite{Dong:2013qoa} finds the following entropy formula:
\begin{equation}
S=\int_W \sqrt{\gamma}\,d^{D-2}\sigma\,\left.\left(\delta^{(1)}R_{\mu\nu\rho\sigma}\frac{\delta \mathcal{L}}{\delta R_{\mu\nu\rho\sigma}}+\sum_{\alpha}\frac{\delta^{(2)}R_{\mu\nu\rho\sigma\tau\pi\xi\zeta}}{1+q_\alpha}\left.\frac{\partial^2 \mathcal{L}}{\partial R_{\mu\nu\rho\sigma}\,\partial R_{\tau\pi\xi\zeta}}\right|_\alpha\right)\right|_{W}\,,
\label{DongEntropy}\end{equation}
with $\delta^{(1)}R_{\mu\nu\rho\sigma}$ and $\delta^{(2)}R_{\mu\nu\rho\sigma\tau\pi\xi\zeta}$ as in \eqref{delta1R} and \eqref{delta2R}. The $\alpha$ in $|_{\alpha}$ in \eqref{DongEntropy} is a dummy variable that takes different values for terms with a different number of parallel/orthogonal projections of the Riemann tensor to $W$ in the $\partial^2 \mathcal{L}/\partial \textrm{Riem}^2$ term. $q_\alpha$ is a sum of numbers that can be $0$, $1/2$ or $1$ depending on the number and kind of projections picked by $\alpha$.\footnote{The case studied in the main text sets $q_\alpha=0$, and is therefore valid for theories in which $\partial^2 \mathcal{L}/\partial \textrm{Riem}^2$ is independent of the curvature, namely curvature squared gravity.}

To be specific, one should add to $q_\alpha$ a factor of $1/2$ for each term of $K_{ab}{}^i$ and $R_{aijk}$ in $\partial^2 \mathcal{L}/\partial \textrm{Riem}^2$; a factor of $1$ for each factor of $\dot{K}_{ab\{ij\}}$; and $0$ otherwise. These quantities can be defined by working out the expansion of the metric around $W$ in \eqref{EmbeddingLeading} to second order in $1/\lambda$:
\beq
\begin{split}
ds^2=\left(\gamma_{ab}(\sigma/\lambda)-\frac{2}{\lambda}K_{abi}(\sigma/\lambda){} x^i-\frac{1}{\lambda^2}\dot{K}_{abij}x^ix^j\right)d\sigma^a d\sigma^b-\frac{1}{\lambda^2}F_{abij} x^j\sigma^b\,d\sigma^a dx^i&\\
-\frac{4}{3\lambda^2}R_{aijk}\,x^i x^k d\sigma^a dx^j+\left(\delta_{ik}-\frac{1}{3\lambda^2}R_{ijkl}x^j x^l\right)dx^i dx^k+O(1/\lambda^3)&\,,
\end{split}\label{coords}\eeq
where we kept normal coordinates in $W$
\beq
\gamma_{ab}(\sigma/\lambda)=\delta_{ab}-\frac{1}{3\lambda^2}\, {}^{(W)} R_{acbd}\sigma^{c}\sigma^{d}+\dots\,.
\eeq

As in eq.~\eqref{EmbeddingLeading}, we have also taken normal coordinates in the space transverse to $W$: $x^i$, $i=1,2$. These are related to those in \eqref{EmbeddingLeading} by $x^1=r\cos\theta$ and $x^2=r\sin\theta$. $x^i$ affinely parametrize geodesics orthogonal to the worldvolume (at constant $\sigma^a$). As in \eqref{EmbeddingNoK}, $\lambda$ is a book-keeping parameter of the curvature lengthscale. Eq.~\eqref{coords} represents a general background metric in coordinates adapted to the most general embedding of $W$, to $O(1/\lambda^{3})$.

The curvatures satisfy  $K_{ab}{}^i=K_{(ab)}{}^i$, $F_{ab}{}^{ij}=F_{[ab]}{}^{[ij]}$ and $\dot{K}_{ab}{}^{ij}=\dot{K}_{(ab)}{}^{(ij)}$. $F_{ab}{}^{ij}$ is a field strength for the $SO(2)$ rotations of the orthogonal frame on $W$, which are a gauge symmetry on $W$.
The Riemann tensor of the background at the origin of $W$ is $R_{\mu\nu\rho\sigma}$, and the following Gauss-Codacci(-like) relations  are satisfied:\footnote{See for instance \cite{Carter:1992vb}. The notation is such that, \emph{e.g.} $R_{abcd}$ is the projection of $R_{\mu\nu\rho\sigma}$ onto $W$ and ${}^{(W)}R_{abcd}$ is the intrinsic Riemann tensor in $W$, with metric $\gamma_{ab}$.}
\begin{align}
R_{abcd}=&{}^{(W)}R_{abcd}-2K_{a[c}{}^iK_{d]bi}\,,\\
R_{abci}=&-2\partial_{[a}K_{b]ci}\,,\label{Rabci}\\
R_{ab}{}^{ij}=&F_{ab}{}^{ij}-2K_{a}{}^{c[i}K_{cb}{}^{j]}\,,\\
R_{a}{}^{i}{}_{b}{}^{j}=&\frac{1}{2}F_{ab}{}^{ij}+K_{a}{}^{cj}K_{cb}{}^{i}+\dot{K}_{ab}{}^{ij}\,,
\end{align}
where it is not necessary to antisymmetrise the $ab$ indices of the $K^2$ term in the third equation because of the speciality of codimension $2$.

To understand the origin of the $\alpha$ prescription in \eqref{DongEntropy}, start by noticing that to introduce a regular conical excess in the metric \eqref{coords}, as was done in \eqref{GeneralRegulatedExp}, one needs to add the regulating factors of eq.~\eqref{GeneralRegulatedExp} in front of the extrinsic curvature terms,
\beq
K_{abi}\,x^i\rightarrow \left(\frac{r}{\Lambda}\right)^{(n-1)B(r^2/\Lambda^2)}K_{abi}\,x^i\,,
\label{ChangeExtrinsic}\eeq
as well as new factors in front of the $ij-$traceless part of $\dot{K}_{abij}$, $\dot{K}_{ab\{ij\}}\equiv \dot{K}_{abij}-1/2\,\delta_{ij}\dot{K}_{abk}{}^k$:
\beq
\dot{K}_{ab\{ij\}}\,x^i x^j \rightarrow \left(\frac{r}{\Lambda}\right)^{2(n-1)C(r^2/\Lambda^2)}\dot{K}_{ab\{ij\}}\,x^i x^j\,,
\label{RegularKdot}\eeq
and $R_{aijk}$,
\beq
R_{aijk} x^i x^k dx^j= \frac{\epsilon^{kj}}{2}R_{aijk}x^i r^2d\theta
\rightarrow \left(\frac{r}{\Lambda}\right)^{(n-1)G(x^2/\Lambda^2)} \frac{\epsilon^{kj}}{2}R_{aijk}x^i r^2d\theta\,.
\eeq
$C(r^2/\Lambda^2)$ and $G(r^2/\Lambda^2)$ have the same properties as $A(r^2/\Lambda^2)$.\footnote{Analogous caveats to the ambiguity in $B(0)$ apply here. We fix them again chosing minimality: $C(0)=G(0)=1$.} These changes are necessary for regularity of \eqref{coords} after the introduction of the `regular' conical excess, as in \eqref{ConicalCartesianCoords}. The remaining terms in \eqref{coords} do not need further regulation because they do not contain dependence in $\theta$.

Let us now review the origin of the terms that one needs to sum in the $q_\alpha$ \cite{Dong:2013qoa}. This has to do with extra $y$ dependence in the $y$ integrals. First of all, let us recall that the strategy is to compute on manifolds with regulated conical excesses, and then send the regulating cutoff $\Lambda\rightarrow 0$. As an example of a term that can show up in the regulated geometries, consider a theory in which $\partial^2\mathcal{L}/\partial\textrm{Riem}^2$ contains a term with a $t$-th power of the extrinsic curvature of $W$. This may generate a contribution to the entropy, \emph{e.g.}:
\beq
\lim_{n\rightarrow 1}\partial_n\lim_{\Lambda\rightarrow 0}\int \sqrt{g}\,d^D x\, \left({}^{(R)}K_{efi}\right)^{t}\, \delta R_{\hat{\sigma}^{a}\hat{r}\hat{\sigma}^{b}\hat{r}}\, \delta R_{\hat{\sigma}^{c}\hat{r}\hat{\sigma}^{d}\hat{r}}\,,
\label{ExampleKt}\eeq
where ${}^{(R)}K_{abi}$ is defined below and the $\delta\textrm{Riem}$ are as in \eqref{Rarbr}. One generally expects contributions from all allowed combinations \eqref{Rarbr}-\eqref{Rathetabtheta}. For the same reasons as in the main text, one needs to evaluate first the integral \eqref{ExampleKt} in the manifold with the regulated conical excess for integer $n\geq2$, send the cutoff $\Lambda$ to zero,  analytically continue the result to $n\sim 1$ and take the derivative and the limit $n\rightarrow 1$.

For $n\neq1$,  $K_{abi}$ in the regulated geometry has changed to a regulated value, ${}^{(R)}K_{abi}$, that can be read from \eqref{ChangeExtrinsic}
\beq
{}^{(R)}K_{abi}=\left(\frac{r}{\Lambda}\right)^{(n-1)B(r^2/\Lambda^2)}\left(K_{abi}+O(n-1)\right)\,.
\label{RegulatingK}\eeq 
By the arguments above eq.~\eqref{ListIntegrals}, this modifies the integrals on $y$ of the type \eqref{Examplento1} to integrals which are effectively of the type:
\beq
\lim_{n\rightarrow 1}\partial_n\int_0^{\infty}dy\, (n-1)^2y^{2n-3+t(n-1)}A(y^2)=\frac{1}{2}\frac{1}{1+t/2}\,,
\label{intyt}\eeq
where $t$ is power of the extrinsic curvature in \eqref{ExampleKt}. The dependence in $t(n-1)$ in the integrand comes from the factor with $y^{(n-1)B(y^2/\Lambda^2)}$ in the regulated extrinsic curvature \eqref{RegulatingK}, $t$ times.

An analogous analysis holds for $R_{aijk}$. For $\dot{K}_{ab\{ij\}}$, the same is true with the substitution $t\rightarrow 2t$, see eq.~\eqref{RegularKdot}. These contributions are clearly additive in the exponent of $y$ in eq.~\eqref{intyt}, and therefore in the rhs of eq.~\eqref{intyt}. It follows that $q_\alpha$ receives the contributions stated at the beginning of the appendix:\footnote{Some factors of $K_{ab}{}^i$ could appear derived, as in $R_{abci}$, \emph{cf.} eq.~\eqref{Rabci}. We need to weight them with the same factor of $1/2$.}
\beq
q_{\alpha}=\frac{1}{2}(\textrm{\# of $K_{abi}\,$s})+\frac{1}{2}(\textrm{\# of $R_{aijk}\,$s})+(\textrm{\# of $\dot{K}_{ab\{ij\}}\,$s})\,.
\label{qalphas}\eeq

Considering general theories of gravity \eqref{GeneralAction}, with derivatives of the Riemann allowed in the action, will probably mildly change the prescription for $q_\alpha$ \eqref{qalphas} \cite{Dong:2013qoa}. We expect new contributions coming from the regularization of the terms represented by $O(1/\lambda^3)$ in eq.~\eqref{coords}, that contain derivatives of the Riemann around the origin of $W$.

The prescription \eqref{DongEntropy} with \eqref{qalphas} reduces to \eqref{NewEntropy} in the case of curvature squared gravity because there $\partial^2\mathcal{L}/\partial{\textrm{Riem}}^2$ is a function of the metric only. Then, there are no contributions of $K_{abi}$, $R_{aijk}$ or $\dot{K}_{ab\{ij\}}$, and $q_\alpha=0$.

\end{document}